\documentclass{ifacconf}

\usepackage{graphicx}      %

\makeatletter
\let\old@ssect\@ssect %
\makeatother

\usepackage{natbib}        %

\usepackage{amsmath,amssymb} 

\usepackage{color}
\usepackage[usenames]{xcolor}
\usepackage{tikz}
\tikzset{beameritem/.style={circle,inner sep=0,minimum size=2ex,text=enumerate item.bg,fill=enumerate item.fg,font=\footnotesize}}
\usepackage{pgfplots} %
\pgfplotsset{compat=newest}
\usepackage{pgfpages}
\usepackage{pdfpages}
\usepackage{siunitx}
\usepackage{url}

\usetikzlibrary{backgrounds}
\usetikzlibrary{shapes}

  \newtheorem{asum}[thm]{Assumption}
    \newtheorem{defi}[thm]{Definition}
      \newtheorem{rema}[thm]{Remark}
 \newcommand{\norm}[1]{\left\lVert#1\right\rVert}
 \newcommand{\abs}[1]{\left|#1\right|} 
 
 \newcommand{\walkl}{\text{\normalfont walk}}
  \newcommand{\epsilonm}{\epsilon}
   \newcommand{\naive}{maximum number of dropouts }
 \newcommand{\binomangle}[2]{\genfrac{\langle}{\rangle}{0pt}{}{#1}{#2}}
\newcommand{\arxivchange}[2]{#2}
 \newcommand{\change}[2]{{#2}}
 \newcommand{\lvar}{\Lambda}

\makeatletter
\newcommand*{\defeq}{\mathrel{\rlap{%
			\raisebox{0.3ex}{$\m@th\cdot$}}%
		\raisebox{-0.3ex}{$\m@th\cdot$}}%
	=}
\makeatother

\begin{document}
	\begin{frontmatter}
		
		\title{\change{Stabilization of}{Stability Analysis} for Nonlinear Weakly Hard Real-Time Control Systems} 

		\thanks[footnoteinfo]{\arxivchange{}{© 2020 the authors. This work has been accepted to IFAC for publication under a Creative Commons Licence CC-BY-NC-ND.} Funded by
			the Deutsche Forschungsgemeinschaft (DFG, German Research Foundation) -
			285825138; 390740016.}
		\author{Michael Hertneck,} 
		\author{Steffen Linsenmayer,} 
		\author{Frank Allg\"ower}
		
		\address{Institute for Systems Theory and Automatic Control, \\			
			 University of Stuttgart, Stuttgart, Germany. \{hertneck,linsenmayer,allgower\}@ist.uni-stuttgart.de.}
		\begin{abstract}                %
This paper considers the \change{stabilization}{stability analysis} for nonlinear sampled-data systems with failures in the feedback loop. The failures are caused by shared resources, 
  and modeled by a weakly hard real-time (WHRT) dropout description. The WHRT dropout description restricts the considered dropout sequences with a non-probabilistic, window based constraint, that originates from schedulability  analysis.
  The proposed approach is based on the emulation of a controller for the nonlinear sampled-data system from a  continuous-time feedback. The emulation technique is extended and combined with non-monotonic Lyapunov functions and a graph description for the WHRT constraints to guarantee asymptotic stability. The effectiveness of the proposed approach is illustrated with a numerical example from literature. \arxivchange{}{This paper is the accepted version of \cite{hertneck20stability}, containing also proofs of the main results.}
		\end{abstract}		
		\begin{keyword}
			Networked embedded control systems, control over networks, control under communication constraints
		\end{keyword}
		
	\end{frontmatter}

\section{Introduction}
In many modern control applications as e.g. in networked or embedded control systems, it may by unavoidable that the control system runs in open-loop from time to time due to failures in the feedback channel. Such failures can be caused e.g. by an unreliable communication system connecting sensors, controllers and actuators (\cite{kauer2014fault}), or since the control input is calculated on a microprocessor that runs many different tasks in parallel, such that the on-time calculation of the control input cannot always be guaranteed (\cite{bernat2001weakly,aarzen2005implementation}). Nevertheless, it is important to guarantee stability of the control system despite the unreliable feedback channel.
For random loss processes, this problem has been thoroughly studied, cf. \cite{schenato2007foundations} for a survey. However, drawbacks of a random dropout description are that stability can only be guaranteed in a mean-square sense and that the random dropout description does often not yield a precise characterization of the loss process, cf. \cite{blind2015towards}: If there are at most 10\% of failures, then this might e.g. mean that there is one failure within 10 tries or that there are 100 consecutive failures after 900 consecutive successes.

An alternative approach to describe the loss process is to determine an upper bound on the successive number of failures in the feedback loop. This maximum number of dropouts approach was studied e.g. in \cite{xiong2007stabilization,kauer2014fault} for linear systems. Moreover, results that guarantee stability, if the time between two arriving control inputs is bounded, as e.g. the emulation technique for nonlinear systems (cf. \cite{carnevale2007lyapunov,nesic2009explicit}), can be used to guarantee stability despite a bounded number of successive failures in the feedback loop. This works by choosing the sampling period smaller or equal than the maximum admissible sampling period (MASP) for the respective approach divided by the maximum number of successive failures plus 1. 

However, to avoid conservatism, it is advantageous to use preferably much information on the loss process for the controller design instead of only employing knowledge about the maximum number of successive failures in the feedback loop. 
 A conceptual framework for a description of the loss process is given by weakly hard real-time (WHRT) constraints (\cite{bernat2001weakly}). WHRT constraints provide guarantees for time windows of fixed size and include a large class of scheduling constraints as e.g. $(m,k)$-firmness from \cite{hamdaoui1995dynamic}. 
 
 In \cite{horssen2016performance}, the problem of controlling a linear system with an $(m,k)$-firmness dropout description has been investigated.
  For linear systems with general WHRT constraints as dropout description, a sufficient stability condition was proposed in \cite{blind2015towards} and controller design methods have been presented in \cite{linsenmayer2017stabilization,linsenmayer2019linear}. \change{}{Conditions for observability and controllability for linear systems with a similar dropout model have been presented in \cite{jungers2018observability}.}  Whereas stability analysis and controller design methods for linear systems with failures in the feedback loop subject to WHRT constraints are thus well established, there are no comparable results for nonlinear systems available in literature.
In this paper, we propose an approach for the stability analysis of nonlinear sampled-data systems with failures in the feedback loop that are described by a weakly hard real-time dropout description. The proposed approach is based on an extension of the emulation  approach from \cite{nesic2009explicit}. A controller that is stabilizing for continuous-time feedback is emulated and an upper bound on a Lyapunov function candidate for the nonlinear sampled-data system is derived. This bound is combined with the concepts of non-monotonic Lyapunov functions (cf. \cite{michel2015stability}) and WHRT graphs (cf. \cite{linsenmayer2019linear}) to obtain a sufficient condition for asymptotic stability. The proposed approach can be used for a wide class of nonlinear systems and is less conservative than considering only the maximum number of dropouts.  We illustrate the benefits of the proposed approach with a numerical example from literature that relates the proposed approach to the maximum number of dropouts approach.

The remainder of this paper is structured as follows. First, we specify in Section~\ref{sec_setup} the considered setup and recap some results from literature. Then we discuss in Section~\ref{sec_preliminary}  the controller emulation technique from \cite{nesic2009explicit} and modify it to our setup. In Section~\ref{sec_main}, we present  sufficient conditions for asymptotic stability of the WHRT control system. Section~\ref{sec_ex} contains \change{a numerical}{an} example to illustrate the proposed approach. A conclusion is given in Section~\ref{sec_conclusion}. \arxivchange{\change{}{Some proofs are omitted due to spacial limitations. They can be found in the preprint \cite{hertneck20arxivb}.}}{}

\subsubsection*{Notation}
The positive, respectively nonnegative, real numbers are denoted by $\mathbb{R}_{>0}$,  respectively  $\mathbb{R}_{\geq 0} = \mathbb{R}_{> 0} \cup \{0\} $. The positive  natural numbers are denoted by $\mathbb{N}$, and $\mathbb{N}_0\defeq\mathbb{N}\cup  \left\lbrace 0 \right\rbrace $. A continuous function $\alpha: \mathbb{R}_{\geq 0} \rightarrow \mathbb{R}_{\geq 0}$ is a class $ \mathcal{K}$ function if it is strictly increasing and $\alpha(0) = 0$. It is a class $\mathcal{K}_\infty$ function if it is of class $\mathcal{K}$ and it is unbounded. The notation $t^-$ is used as $t^- \defeq \lim\limits_{s<t,s\rightarrow t} s$. A continuous function $V:\mathbb{R}^n \rightarrow \mathbb{R}$ is positive definite if $V(0) = 0$ and $V(x)>0$ for all  $x\backslash \left\lbrace 0 \right\rbrace$.  
\section{Setup}
\label{sec_setup}
\arxivchange{}{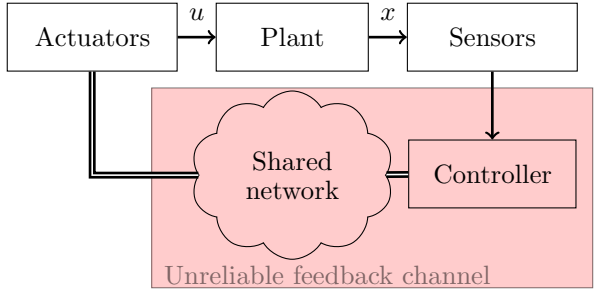
\begin{figure}[tb]
	\centering
\begin{tikzpicture}
	\usetikzlibrary{positioning}
	\usetikzlibrary{backgrounds}
	\usetikzlibrary{shapes,arrows,chains} 
	\usetikzlibrary{patterns}
	
	\pgfdeclarepatternformonly{soft horizontal lines}{\pgfpointorigin}{\pgfqpoint{100pt}{1pt}}{\pgfqpoint{100pt}{3pt}}%
	{
		\pgfsetstrokeopacity{0.3}
		\pgfsetlinewidth{0.1pt}
		\pgfpathmoveto{\pgfqpoint{0pt}{0.5pt}}
		\pgfpathlineto{\pgfqpoint{100pt}{0.5pt}}
		\pgfusepath{stroke}
	}

 		\node[rectangle,inner sep = 0pt,draw, text width=2 cm,align=center, minimum height= 0.9cm] (node1) {Plant};
		\node[draw,right  = 0.5cm  of node1, text width=2 cm,align=center, minimum height= 0.9cm] (node3) {Sensors};
		{\node[draw,left  = 0.5cm  of node1, text width=2cm,align=center, minimum height= 0.9cm] (node5) {Actuators};}

 		\node[left = 1 cm of node5](helpnode){};
		\node[below = 1.75 cm of helpnode](helpnode2){};

 		\begin{scope}[on background layer]
 				\node[draw,fill = red!40,right = 2.9 cm of helpnode2,fill,rectangle,semitransparent,minimum width=5.8 cm,minimum height=2.65cm](box_network){};
 				\node[below = -0.3 cm of box_network](helpnode3){};
 				\node[left = -1.8 cm of helpnode3,rectangle,semitransparent](box_network2){Unreliable feedback channel};
		
 		\end{scope}
 		
 		\node[draw,below  = 0.9cm  of node3, rectangle,inner sep = 0pt, text width=2.2 cm,align=center, minimum height= 0.9cm] (node7) {Controller};
 		\node [cloud, draw, below  = .25 cm  of node1, cloud puffs=10,cloud ignores aspect,cloud puff arc=120, aspect=2, inner ysep=1em,text width = 1.5cm,minimum height = 0.6 cm, align = center] (node6) {Shared \\ network};
 		
 		\draw [->,line width = 1pt] (node1) -- (node3);
 		\draw [->,line width = 1pt] (node5) -- (node1);
		{\draw [-,line width = 1pt,double] (node6) -| (node5);}
 		{\draw [-,line width = 1pt,double] (node6) -- (node7);}
 		\draw [->,line width = 1pt] (node3) -- (node7);
 		\path [line width = 1pt] (node1) -- node [text width=1.5cm,midway,above = 0.25em,align = center] {$x$} (node3);
 		\path [line width = 1pt] (node5) -- node [text width=1.5cm,midway,above = 0.25em,align = center] {$u$} (node1);
 		\path [line width = 1pt] (node7) -- node [text width=1.5cm,midway,left = 0.25em,align = center] {} (node5);
 \end{tikzpicture}
	\vspace{-0.5cm}
	\caption{Sketch of the considered setup.}
	\label{fig_setup}
\end{figure}}{}
In this section, we specify the setup considered in this paper, introduce the notation of WHRT constraints and non-monotonic Lyapunov functions and formalize the control objective of the paper.
\subsection{Control System}
\arxivchange{}{A sketch of the studied setup is given in Figure~\ref{fig_setup}.} We consider a nonlinear, time-invariant plant
\begin{equation}
\label{eq_sys_cont}
\dot{x} = f_p(x,u)
\end{equation}
with a continuously differentiable vector valued function $f_p:\mathbb{R}^{n_x}\times\mathbb{R}^{n_u} \rightarrow \mathbb{R}^{n_x}$ satisfying $f_p(0,0) = 0$, %
the system state $ x(t)\in \mathbb{R}^{n_x}$ with initial condition $x(0) = x_0$ and the input $u(t)\in\mathbb{R}^{n_u}$ that is applied by the actuator.

The system is sampled periodically  with a fixed sampling period $h$ to be specified later. Thus, at each discrete time instant satisfying $t = kh$ for some $k\in\mathbb{N}_0$, a new sample of the system state\change{arrives at the controller}{is sent to the controller}.\change{The controller generates a control input $\hat{u}$ at each sampling time by}{When the controller receives a sample of the state, then a control input $\hat{u}$ is generated as}
\change{\begin{equation}
\label{eq_cont_fb}
\hat{u}(kh) = \kappa(x(kh))
\end{equation}}{$\hat{u}(kh) = \kappa(x(kh))$}
with the nonlinear feedback law $\kappa:\mathbb{R}^{n_x}\rightarrow\mathbb{R}^{n_u}$. 
We assume that $\kappa$ asymptotically stabilizes the continuous-time system, \change{which can be formalized as the following assumption}{according to the following assumption}.
\begin{asum}
	\label{as_cont_fb}
	There is a locally Lipschitz, positive definite function $V:\mathbb{R}^{n_x} \rightarrow \mathbb{R}$ satisfying for all $x\in \mathbb{R}^{n_x}\backslash \left\lbrace 0 \right\rbrace$
	\begin{equation}
	\underline{\alpha}_V(\norm{x}) \leq V(x) \leq \overline{\alpha}_V(\norm{x})
	\end{equation}
	for $\underline{\alpha}_V,\overline{\alpha}_V \in \mathcal{K}_\infty$ and
	\begin{equation}
		\left\langle\nabla V(x),f_p(x,\kappa(x)) \right\rangle <0.
	\end{equation} 
	
\end{asum}
From time to time, \change{}{transmissions of the sampled state to the controller,} the calculation of the control input, or its transmission to the actuator, fail due to higher priority tasks on the microprocessor or due to imperfections of the communication system. 
The time instants, when a new input is received at the actuator are given by the infinite sequence $(\tau_z)_{z \in \mathbb{N}_0}$  and define a discrete set 
\begin{equation}
\label{eq_T_def}
\mathcal{T} \defeq \left\lbrace \tau_0, \tau_1,\tau_2,\dots \right\rbrace.
\end{equation}
For simplicity, we assume subsequently that the control law for the initial state is received successfully at the actuator at the first sampling time and have thus  $\tau_0 = 0$.
Moreover, we assume that there is no transmission delay, i.e., for each $z\in\mathbb{N}_0$, $\tau_z = kh$ for some $k\in \mathbb{N}_0$.
With $\mathcal{T}$, we can furthermore specify an infinite binary sequence $\nu\defeq(\nu_k)_{k\in\mathbb{N}_0}$ that describes, if a new input is received at sampling instants. We set $\nu_k = 1$ if $kh \in \mathcal{T}$, i.e., if a new input is received at sampling instant $k$, and $\nu_k = 0$ if $kh \notin \mathcal{T}$, i.e., if no new input arrives.  
Between times from $\mathcal{T}$, the actuator holds \change{the system input as}{}the last received feedback, i.e., 
\begin{equation}
\label{eq_fb_zoh}
	u(t) = \hat{u}(\tau_z)=  \kappa(x(\tau_z)), \tau_z \leq t < \tau_{z+1}
\end{equation}
for all $z\in\mathbb{N}_0$. We define the sampling error as 
\begin{equation}
e(t) = x(\tau_z) - x(t), \tau_z\leq t < \tau_{z+1} 
\end{equation}
for all $z\in\mathbb{N}_0$.
We can thus model the closed-loop system composed of \eqref{eq_sys_cont} with input defined by \eqref{eq_fb_zoh} and a \change{given}{}dropout sequence $\nu$ 
as the discontinuous dynamical system (DDS) 
\begin{align}
&\left.\begin{aligned}
\nonumber \dot{x}(t)&= f(x(t),e(t))\defeq  f_p(x,\kappa(x+e)) \\
\dot{e}(t) &= g(x(t),e(t)) \defeq  -f_p(x,\kappa(x+e)) 
\end{aligned}\right\}, \quad \tau_z \leq t < \tau_{z+1},\\
\label{eq_def_dds}
&\left.\begin{aligned}
x(t)&= x(t^-) \\
e(t) &= 0 
\end{aligned}\right\}, \hspace{1.3 cm}   t = \tau_{z+1},~\forall z \in \mathbb{N}_0
\end{align}

with $x(0) = x_0$ and $e(0) = 0$.
A trajectory that satisfies \eqref{eq_def_dds} for all $t \geq 0$ exists, if a unique solution to
\begin{align}
	\begin{split}
		\dot{\tilde{x}}(t) =& f(\tilde{x}(t),\tilde{e}(t))\\
		\dot{\tilde{e}}(t) =& g(\tilde{x}(t),\tilde{e}(t)) \label{eq_aux_sys}
	\end{split}
\end{align}
  exists for $\tau_z \leq t \leq \tau_{z+1} $ for all $z \in \mathbb{N}_0$ and arbitrary $\tilde{x}(\tau_z)$ for $\tilde{e}(\tau_z) = 0$. We assume from now on that such \change{a}{}solution\change{}{s} exist\change{s}{} and justify later why this assumption is reasonable.

\subsection{Weakly Hard Real-Time Constraints}
\label{subsec_whrt}
As in the recent works \cite{blind2015towards,linsenmayer2017stabilization,linsenmayer2019linear}, we use WHRT constraints to model the loss process, i.e., to specify the sequence $\nu$ from the previous subsection. Thus, we shall also use the following definitions that are taken from \cite{bernat2001weakly} and \cite{blind2015towards}.

\begin{defi}(cf. \cite[Def. 1]{blind2015towards})
	A constraint $\eta$ is a function that maps an infinite binary sequence $\nu$ to the Boolean values {\tt true} and {\tt false}, i.e., $\eta:\left\lbrace 0,1\right\rbrace^\infty \rightarrow \left\lbrace{\tt true}, {\tt false} \right\rbrace$. We say that a sequence $\nu$ satisfies a constraint $\eta$, denoted by $\nu \vdash \eta$, when $\eta(\nu) = {\tt true}$.
\end{defi}
\begin{defi} (cf. \cite[Def. 3]{bernat2001weakly}
	A transmission sequence $\nu$ ``meets any $n$ in $m$ deadlines" ($m\geq1, 1 \leq n \leq m$) and it is denoted by $\nu \vdash \binom{n}{m}$ if, in any window of $m$ consecutive transmissions, there are at least $n$ transmissions in any order, that are successful.  
	\end{defi}
\begin{defi} (cf. \cite[Def. 4]{bernat2001weakly}
	A transmission sequence $\nu$ ``meets row $n$ in $m$ deadlines" ($m\geq1, 1 \leq n \leq m$) and it is denoted by $\nu \vdash \binomangle{n}{m}$ if,  in any window of $m$ consecutive transmissions, there are at least $n$ consecutive transmissions, that are successful.  
\end{defi}
\begin{defi} (cf. \cite[Def. 6]{bernat2001weakly}
	A transmission sequence $\nu$ ``misses less than row $n$ in $m$ deadlines" ($m\geq1, 1 \leq n \leq m$) and  it is denoted by $\nu \vdash \overline{\binomangle{n}{m}}$ if, in any window of $m$ consecutive transmissions, it is never the case that $n$ consecutive transmissions are unsuccessful. 
\end{defi}
\begin{defi}
	(cf. \cite[Def. 10]{bernat2001weakly}): Given two constraints, $\eta$ and $\eta^\prime$, we say that $\eta^\prime$ is harder than $\eta$ ($\eta$ is easier than $\eta^\prime$), denoted by $\eta^\prime \preceq \eta$ if all sequences that satisfy $\eta^\prime$ also satisfy $\eta$.
\end{defi}

\begin{defi}
	The maximum number of consecutive lost transmissions, that may be contained in a sequence $\nu$, such that $\nu$ can still satisfy $\eta$, is denoted by $w(\eta)$.
\end{defi}

Henceforth, we use the graph representation for WHRT constraints, that has been proposed in \cite{linsenmayer2017stabilization} and extended in \cite{linsenmayer2019linear}, to capture all possible transmission sequences that satisfy a constraint $\eta$. A labeled directed graph is described by a tuple $\mathcal{G} = (\mathcal{V},\mathcal{E})$, where $\mathcal{V} = \left\lbrace v_1,\dots,v_{n_\mathcal{V}} \right\rbrace$  is the set of nodes
 and $\mathcal{E} = \left\lbrace e_1,\dots,e_{n_\mathcal{E}} \right\rbrace$ is the set of edges.  If there exists an edge from node $v_{i_p}$ to node $v_{j_p}$ with weight $l_p$, then $e_p = (i_p,j_p,l_p)$ is contained in $\mathcal{E}$. A walk $P$ for a graph $\mathcal{G}$ is a sequence of edges of $\mathcal{G}$ and can be described by a (possibly infinite)  index sequence $\mathbb{I}(P) = \left( \mathbb{I}_1(P),\mathbb{I}_{2}(P),\dots\right) $ of edge indices\change{}{, where $\mathbb{I}_k(P) \in \mathbb{N}$ for all $k \in \mathbb{N}$}. Thus, the sequence of edges of $P$ is described by  $ \left( e_{\mathbb{I}_1(P)}, e_{\mathbb{I}_2(P), \dots} \right)$.  If the number of elements in $P$ is finite, then we denote this number by $n_P$. 
Note, that an edge may be contained multiple times in a walk.  The cost of a walk is the sum of the edge weights of the walk.

\begin{figure}[tb]
	\centering
	\scalebox{0.65}{
		\begin{tikzpicture}[->,shorten >=0pt,auto,node distance=2.8cm,
		thick,main node/.style={circle,draw,font=\Large\bfseries}]
		\node[main node] (A) {$v_1$};
		\node[main node] (B) [above right of=A] {$v_2$};
		\node[main node] (C) [right of=A] {$v_3$};
		\path
		(A) edge [loop left] node {\textcolor{black}{$e_3 (l_3 = 1)$}} (A)
		edge node [below, right = 1]{$e_5 (l_5 = 2)$} (B)
		edge [bend right=10] node [below]{\textcolor{black}{$e_2 (l_2 = 3)$}} (C)
		(B) edge [bend right=10] node [above, left = 1]{\textcolor{black}{$e_4 (l_4 = 1)$}} (A)
		(C) edge  node[above] {\textcolor{black}{$e_1 (l_1 = 1)$}} (A);        
		\end{tikzpicture}}
	\caption{WHRT graph for $\binomangle{2}{5}$.}
	\label{fig_whrt}
\end{figure}

 In order to describe a constraint $\eta$ by a graph $\mathcal{G}_\eta$, we use the definition of WHRT graphs from \cite{linsenmayer2019linear}, that can be given for our setup as follows.
 
 \begin{defi}
 	(cf. \cite{linsenmayer2019linear}, Def. 9) A labeled directed graph $\mathcal{G}_\eta$ is a WHRT graph for a given weakly hard real-time constraint $\eta$, if for all sequences $\nu \vdash \eta$, there exists an infinite walk $P$ for $\mathcal{G}_\eta$, such that 
 	\begin{equation}
 		\nu = (1 \underbrace{0 \dots 0}_{l_{\mathbb{I}_1(P)}-1~\text{\normalfont times}} 1 \underbrace{0 \dots 0}_{l_{\mathbb{I}_2(P)}-1~\text{\normalfont times}} 1  \dots). \label{eq_conc_edge}
 	\end{equation}
 \end{defi}
 
 Thus, the edge labels of a WHRT graph $\mathcal{G}_\eta$ do represent the number of sampling periods between two successfully received inputs, and we can generate all dropout sequences that satisfy a constraint $\eta$ by concatenating edges of $\mathcal{G}_\eta$ in a similar fashion as in \eqref{eq_conc_edge}. 
 For a constraint $\eta$ of type $\binom{n}{m}$, a WHRT graph $\mathcal{G}_\eta$, can be constructed automatically\footnote{Matlab code can be found at \url{https://www.ist.uni-stuttgart.de/institute/team/pdf/SL/WhrtGraphMatlab.zip}.} using Algorithm~1 from \cite{linsenmayer2019linear}. For the constraint types ${\binomangle{n}{m}}$ and $\overline{\binomangle{n}{m}}$, construction algorithms can easily be derived based on the same main ideas as in Algorithm~1 from \cite{linsenmayer2019linear}.
 The WHRT graph for $\binomangle{2}{5}$ is given as example in Figure~\ref{fig_whrt}.	
For this graph, an exemplary walk $P_e$ can be described by the index sequence $\mathbb{I}(P_e) = (2,1,3,5,4,2)$ that generates the sequence $\nu = (1,0,0,1,1,1,0,1,1,0,0,1)$.  	 We will later use furthermore the following definition.
 	 \begin{defi}
 	 	We denote by $\mathcal{S}(\mathcal{G},c_\walkl)$ the set of all walks of a graph $\mathcal{G}$ with cost larger or equal than $c_\walkl \in \mathbb{N}$, that are such that removing the last edge from the walk makes the cost of the walk smaller than $c_\walkl$. 
 	 \end{defi}
 	 Each walk from $\mathcal{S}(\mathcal{G}_\eta,c_\walkl)$ can be used to create a (finite) binary transmission sequence. Note, that we can generate each infinite sequence $\nu \vdash\eta$ that starts with a successful transmission by appending suitable walks from $\mathcal{S}(\mathcal{G}_\eta,c_\walkl)$ to an infinite walk and concatenating the edge labels of that walk according to \eqref{eq_conc_edge}. 
 	 
\subsection{Non-Monotonic Lyapunov Functions}
Next, we present a sufficient stability condition for the DDS model~\eqref{eq_def_dds},  that can be derived similarly as Theorem~6.4.6 from~\cite{michel2015stability}. 
This condition is based on a non-monotonic Lyapunov function that may increase for some times, as long as one can still guarantee an average decrease, 
and will thus turn out to be useful in our setup with a WHRT dropout description for the feedback loop. It can be formulated as follows.	

\begin{prop}
	\label{prob_michel}
	Observe the DDS given by \eqref{eq_def_dds}. Let $\xi=\left[x^\top,e^\top\right]^\top$. 
	Assume that there is an unbounded discrete subset $\tilde{\mathcal{T}} \subseteq \mathcal{T}$, described by an infinite sequence $(\tilde{\tau}_z)_{z\in\mathbb{N}_0}$ that satisfies 
	\begin{equation}
	\label{eq_non_mon_dec3}
	0 < \underline{h} \leq \tilde{\tau}_{z+1} - \tilde{\tau}_{z} \leq \overline{h},~ \forall z\in\mathbb{N}_0
	\end{equation}
	and $\tilde{\tau}_0 = 0$. Moreover, assume there is a continuous positive definite function $V_n:\mathbb{R}^{2n_x} \rightarrow \mathbb{R},$ satisfying 
		\begin{equation}
		\label{eq_non_mon_dec4}
		\alpha_1 {(\norm{\xi})} \leq V_n(\xi) \leq \alpha_2 {(\norm{\xi})} 	
		\end{equation}
	 such that for  all $z\in \mathbb{N}_0,$
	\begin{equation}
	\label{eq_non_mon_desc1}
	V_n(\xi(\tilde{\tau}_z+r)) \leq \alpha_3(V_n(\xi(\tilde{\tau}_z))), ~ 0 \leq r < \tilde{\tau}_{z+1} -\tilde{\tau}_z
	\end{equation}
	and
	\begin{equation}
	\label{eq_non_mon_desc2}
	\frac{1}{\tilde{\tau}_{z+1} - \tilde{\tau}_z} \left[ V_n(\xi(\tilde{\tau}_{z+1})) - V_n(\xi(\tilde{\tau}_z))\right] \leq -\alpha_4 (V_n(\xi(\tilde{\tau}_z)))
	\end{equation}
	hold with class $\mathcal{K}_\infty$ functions $\alpha_1, \alpha_2$ and class $\mathcal{K}$ functions $\alpha_3, \alpha_4$. Then the equilibrium $\xi = 0$ is globally asymptotically stable for \eqref{eq_def_dds}. 
\end{prop}
	\arxivchange{The proof of Proposition~\ref{prob_michel} can be found in the preprint \cite{hertneck20arxivb}.}{\begin{pf}See Appendix~\ref{append_a}.\end{pf}}
The main difference of Proposition~\ref{prob_michel} in comparison to Theorem~6.4.6 from~\cite{michel2015stability} is that additional jumps of the  DDS~\eqref{eq_def_dds} at times from $\mathcal{T}\backslash \tilde{\mathcal{T}}$ may occur.

 \subsection{Problem Statement}
 	 For a given WHRT constraint $\eta$ and a given sampling period $h$, the goal of this paper is to derive a sufficient stability condition for control systems with unreliable feedback loops, described by the DDS~\eqref{eq_def_dds} (or respectively \eqref{eq_sys_cont} and \eqref{eq_fb_zoh}) with a dropout sequence $\nu\vdash \eta$, that exploits the knowledge about $\nu$ that is described by $\eta$. 

\section{Emulation of the Controller}
\label{sec_preliminary}
In this section, we adapt the emulation technique from \cite{nesic2009explicit} that can be used to obtain stability guarantees for the sampled-data system \eqref{eq_def_dds} with emulated continuous-time controller based on a monotonic Lyapunov function, to the setup with non-monotonic Lyapunov functions considered in this paper. More precisely, we modify the explicit bound on the MASP from \cite{nesic2009explicit} such that we are able to guarantee a decrease of $V(x)$, or at least to  bound the amount of increase of $V(x)$ by a know reference, depending on the time between two arriving inputs. 
  First, we state a basic assumption.
\begin{asum}
	\label{asum_hybrid_lyap}
	There exist a locally Lipschitz function $W:\mathbb{R}^{n_x} \rightarrow \mathbb{R}_{\geq0}$, a continuous function $H:\mathbb{R}^{n_x} \rightarrow \mathbb{R}_{\geq0}$, $L, \gamma \in \mathbb{R}_{>0}$, $\epsilon \in \mathbb{R}$ and $\underline{\alpha}_W$, $\overline{\alpha}_W \in \mathcal{K}_\infty$,  such that for all $e\in\mathbb{R}^{n_x} $ and $x\in\mathbb{R}^{n_x}$,
	\begin{equation}
	\label{eq_w_bound}
	\underline{\alpha}_W(\norm{e}) \leq W(e) \leq \overline{\alpha}_W(\norm{e})
	\end{equation}
	and 
	\begin{equation}
	\left\langle \frac{\partial W^2(e)}{\partial e},g(x,e)\right\rangle \leq 2 W(e) \left(L W(e) + H(x) \right) \label{eq_w_est}
	\end{equation}
	hold. Moreover, for $V(x)$ from Assumption~\ref{as_cont_fb},
	\begin{equation}
	\label{eq_v_desc_hybrid}
	\left\langle \nabla V(x),f(x,e) \right\rangle \leq - \epsilonm V(x) -H^2(x) + \gamma^2 W^2(e)
	\end{equation}
	holds.
\end{asum}

This assumption is  comparable to Assumption~1 in \cite{nesic2009explicit}. There are two noteworthy differences. 

We use  $-\epsilonm V(x)$ instead of $-\varrho(x)$ in \eqref{eq_v_desc_hybrid} and thus require exponential stabilizability of the continuous time system for $\epsilon > 0$, which is also often done in literature, see e.g. Assumption~3 in \cite{postoyan2014tracking} and Assumption 8.1 in \cite{heijmans2019exploring}. A possible relaxation in comparison to those assumptions from literature is that $\epsilonm$ in Assumption~\ref{asum_hybrid_lyap}  not necessarily needs to be positive, which will play an important role in our main result. Note that Assumption~\ref{asum_hybrid_lyap} can hold for a DDS~\eqref{eq_def_dds} simultaneously with different combinations of $\epsilon$, $\gamma$ and $L$. In general, a smaller value for $\epsilon$ will allow us to choose $\gamma$ smaller and vice versa. We will require in our main result at least one parameter set with $\epsilonm > 0$, but also exploit different parameter sets for arbitrary $\epsilon$ when possible.

The second difference is that we have replaced  condition (9) from \cite{nesic2009explicit} by \eqref{eq_w_est}. 
In fact, if $W^2(e)$ is differentiable, which is, e.g., the case for $W(e) = \norm{e}$, then  \eqref{eq_w_est} is essentially the same as condition (9) of Assumption~1 in \cite{nesic2009explicit}.

The technique from \cite{nesic2009explicit} to determine a bound on the the MASP for systems with emulated controller, that satisfy Assumption~\ref{asum_hybrid_lyap} for $\epsilon > 0$, $\gamma$ and $L$, is to chose the maximum time between two arriving inputs smaller than the bound $ T_{\max}$, where	
	\begin{equation}
	T_{\max}(\gamma,\lvar ) \defeq \begin{cases}\vspace{1mm}
	\frac{1}{\lvar r} \mathrm{arctan}(r) & \gamma > \lvar \\ \vspace{1mm}
	\frac{1}{\lvar } & \gamma = \lvar \\
	\frac{1}{\lvar r} \mathrm{arctanh}(r) &\gamma < \lvar 
	\end{cases}
	\end{equation}
with
	
	\begin{equation}
	\label{eq_def_r}
	r\defeq\sqrt{\abs{
			\left(\frac{\gamma}{\lvar }\right)^2-1}}
	\end{equation} 
for $\lvar  = L$. Then we can explicitly state a Lyapunov function that is monotonically decreasing. Note that the value of 	$T_{\max}(\gamma,\lvar )$ increases as $\gamma$ or $\lvar $ decrease.

Our goal is to combine the emulation approach with non-monotonic Lyapunov functions. Doing so, we can even tolerate an increase of the  non-monotonic Lyapunov function, but need to upper-bound the amount of its increase or decrease
  at any time. To find such a bound, we 
\change{investigate}{study} now the \change{influence}{effect} of arbitrary $\epsilon$ and $\lvar $ on the time evolution of $V(x(t))$ and $W(e(t))$, if the time between two arriving inputs is bounded by $T_{\max}(\gamma,\lvar )$ for $\gamma, \epsilon$ and $L$ that satisfy Assumption~\ref{asum_hybrid_lyap}.
For this setup, we can state the following proposition.

\begin{prop}
	\label{prop_hybrid}
	Let Assumption~\ref{asum_hybrid_lyap} hold for some $\gamma, \epsilon, L$. Let $0 < \tau_{z+1} - \tau_z < T_{\max} (\gamma,\lvar )$ for some $z \in \mathbb{N}_0$ and $\lvar  >0$. Then, a unique solution to \eqref{eq_aux_sys} exists for $\tau_z \leq t \leq \tau_{z+1}$. Furthermore, for \eqref{eq_def_dds}, it holds that 
	\begin{align}
	\nonumber &V(x(\tau_{z+1})) \\
	\leq& \exp\left(\max\left\lbrace -\epsilonm,2(L-\lvar ) \right\rbrace (\tau_{z+1} - \tau_z)\right) V(x(\tau_z)). \label{eq_prop_hybrid1}
	\end{align}
	Moreover, 	
	\begin{equation}
	V(x(t)) \leq k_1 V(x(\tau_z)) \label{eq_prop_hybrid2}
	\end{equation}
	and
	\begin{equation}
	\norm{e(t) }\leq \alpha_{w2}(V(x(\tau_z)))\label{eq_prop_hybrid3}
	\end{equation}
	hold for $\tau_z \leq t < \tau_{z+1}$, $k_1 \in \mathbb{R}_{> 0}$ and $\alpha_{w2} \in \mathcal{K}$.
\end{prop}

	\arxivchange{The proof for Proposition~\ref{prop_hybrid} can be found in the preprint \cite{hertneck20arxivb}.}{ \begin{pf} See Appendix~\ref{append_b}. \end{pf}}
Proposition~\ref{prop_hybrid} delivers\change{thus}{} a bound on the increase or decrease of $V(x)$ between two arrival times of inputs that depends on $\max\left\lbrace -\epsilonm,2(L-\lvar ) \right\rbrace $, if the time span between the two arrival times is upper bounded by $T_{\max}(\gamma,\lvar )$. For  $\lvar >L$ and $\epsilon > 0$, we can guarantee a certain amount of decrease for $V(x)$ between both arrival times at the cost of a smaller admissible time span between the arrival times. On the other hand, if  $\lvar <L$ or $\epsilon < 0$, which allows a larger time span between two arriving inputs, then $V(x)$ may increase between the arrival times, but the amount of increase remains bounded. 
\change{We will use this together with the graph description for the WHRT constraints to construct a non-monotonic Lyapunov function and thus to obtain a sufficient condition for asymptotic stability of the origin of the DDS~\eqref{eq_def_dds} in the next section.}{We will combine this with WHRT graphs and non-monotonic Lyapunov functions to derive a sufficient condition for asymptotic stability of the origin of the DDS~\eqref{eq_def_dds} in the next section.}

\begin{rema}
	Proposition~\ref{prop_hybrid} justifies also our assumption, that a unique solution to \eqref{eq_aux_sys} exists for $\tau_z \leq t \leq \tau_{z+1}$ for all $z\in \mathbb{N}_0$, arbitrary $x(\tau_z)$ and $e(\tau_z) = 0$, if $\tau_{z+1} - \tau_z$ is small enough\change{}{, which will be ensured by the choice of $h$ in the next section}. \change{This will be ensured by the choice of $h$ in the next chapter.}{}
\end{rema}

\section{Stability Results}
\label{sec_main}
Now, we present sufficient stability conditions for the DDS~\eqref{eq_def_dds} despite failures in the feedback loop\change{ that are subject to a WHRT constraint}{}.

\subsection{Maximum Number of Dropouts Approach}
A straightforward approach to guarantee stability despite the loss process, that does only take  into account  the worst case number of sampling periods between two arriving inputs, i.e.,  $w(\eta)+1$, is the  \naive approach. Here, the sampling period $h$ has to satisfy $h < \frac{T_{\max}(\gamma,L)}{w(\eta)+1}$, with $\gamma$ and $L$ chosen such, that Assumption~\ref{asum_hybrid_lyap} holds for some $\epsilon > 0$. Then, asymptotic stability can be guaranteed with the standard emulation technique.
 For a DDS \eqref{eq_def_dds}, in general, there are possible multiple choices for $\epsilon>0, \gamma$ and $L$, such that Assumption~\ref{asum_hybrid_lyap} holds. Thus, for the  \naive approach, it is beneficial to seek for the combination that maximizes $T_{\max}(\gamma,L)$ and satisfies Assumption~\ref{asum_hybrid_lyap} with $\epsilon>0$.

\subsection{A  Stability Condition Based on WHRT Constraints}

It can be easily verified that $\binomangle{n}{m} \preceq \overline{\binomangle{m-n+1}{*}}$ and $\binom{n}{m} \preceq \overline{\binomangle{m-n+1}{*}}$ for $*$ larger than $m-n+1$. Observe, that the \naive approach is in principle based on constraints of the type $\overline{\binomangle{w(\eta)+1}{*}}$, where $*$ represents an arbitrary window size larger than $w(\eta)$, instead of the actual constraint type.  
 In other words, if the dropout sequence satisfies constraints of the types $\binomangle{n}{m}$ or $\binom{n}{m}$, then a weaker constraint is taken into account by the \naive approach to verify stability, and not the full knowledge on the original constraint is used.
We propose in this subsection a stability condition, which in contrast exploits additional  knowledge on the dropout sequence for constraints of the types $\binomangle{n}{m}$ and $\binom{n}{m}$.

This sufficient condition is based on the WHRT graph $\mathcal{G}_\eta$, which we can compute with tools from \cite{linsenmayer2019linear} for a WHRT constraint $\eta$. For given $c_\walkl>0$, we can moreover determine $\mathcal{S}(\mathcal{G}_\eta,c_\walkl)$ for $\mathcal{G}_\eta$ with a simple recursion. Then, since a new input is received at the first sampling time, we can generate any sequence $\nu \vdash \eta$ by appending walks from $\mathcal{S}(\mathcal{G}_\eta,c_\walkl)$. 
The next step is to compute an upper bound on $V(x)$ for all transmission sequences that can be generated by the walks from $\mathcal{S}(\mathcal{G}_\eta,c_\walkl)$. If this bound is decreasing for each  such sequence, then we can show asymptotic stability of the origin of the DDS~\eqref{eq_def_dds} with Proposition~\ref{prob_michel} for  $\tilde{\mathcal{T}}$ chosen such, that it contains always the starting times and the end times of each appended walk from $\mathcal{S}(\mathcal{G}_\eta,c_\walkl)$.
\change{The resulting sufficient stability condition is given by the following Theorem.}{As a result, we obtain the following sufficient stability condition.}

\begin{thm}
	\label{theo_stab}
	Consider the closed-loop system~\eqref{eq_def_dds} with a sampling period $h\in \mathbb{R}_{>0}$ and a dropout sequence $\nu \vdash \eta$ for a given WHRT constraint $\eta$. Let $\mathcal{G}_\eta$ be a WHRT graph for $\eta$ and let $c_\walkl \in \mathbb{N}$. Let Assumption~\ref{as_cont_fb} hold. Assume moreover, there are parameters $\left(\gamma_i,L_i,\lvar _i,\epsilonm_i\right)$ for each $i \in \left\lbrace 1,\dots,w(\eta)+1 \right\rbrace$, such that Assumption~\ref{asum_hybrid_lyap} holds for  $\left(\gamma_i,L_i,\epsilonm_i\right)$, 
	and  
	\begin{equation}
		\label{eq_sample_bound}
		ih < T_{\max}(\gamma_i,\lvar _i)
	\end{equation} holds for all $ i \in \left\lbrace 1,\dots,w(\eta)+1 \right\rbrace$. \change{If for}{Assume for} each walk $P \in \mathcal{S}(\mathcal{G}_\eta,c_\walkl)$ with index sequence $\mathbb{I}(P)$, and sequence of edge weights of the walk  $\left( l_{\mathbb{I}_1(P)}, \dots, l_{\mathbb{I}_{n_P(P)}} \right)$, \change{it holds that}{that}
	\begin{equation}
	\label{eq_theo_sum}
		\sum_{\theta = 1}^{n_{P}} hl_{\mathbb{I}_\theta(P)} \max \left\lbrace -\epsilon_{l_{\mathbb{I}_\theta(P)}},2(L_{l_{\mathbb{I}_\theta(P)}}-\lvar _{l_{\mathbb{I}_\theta(P)}})\right\rbrace < 0 \change{,}{}
	\end{equation}
	\change{}{holds.} Then the origin of \eqref{eq_def_dds} is asymptotically stable for all sequences $\nu \vdash \eta'$ with $\eta' \preceq \eta$.
\end{thm}

	\arxivchange{The proof of Theorem~\ref{theo_stab} can be found in the preprint \cite{hertneck20arxivb}.}{\begin{pf}See Appendix~\ref{append_c}.\end{pf}}

\change{\begin{rema}
	In principle, the parameter $c_\walkl$ can be chosen arbitrarily. However, in most cases it is advantageous to chose it equal to the window length of the constraint $\eta$.
\end{rema}}{}
\begin{rema}
	For constraints of type $\overline{\binomangle{n}{m}}$, Theorem~\ref{theo_stab} and the \naive approach are equivalent. For constraints of the types $\binomangle{n}{m}$ and $\binom{n}{m}$, Theorem~\ref{theo_stab} can exploit the additional information to guarantee stability for significantly larger sampling periods than the \naive approach.
\end{rema}
\change{}{The parameter $c_\walkl$ can be chosen arbitrarily. However, in most cases it is advantageous to chose it equal to the window length of the constraint $\eta$.}
	For given parameters $\gamma_i,L_i,\epsilonm_i$ and $\lvar _i$, for all $ i \in$ $ \left\lbrace 1,\dots,w(\eta)+1 \right\rbrace$, condition~\eqref{eq_theo_sum} can be automatically verified. For that, the set $\mathcal{S}(\mathcal{G}_\eta,c_\walkl)$ can be determined by a recursive search through $\mathcal{G}_\eta$. Then, the left hand side of \eqref{eq_theo_sum} can be computed for any walk from  $\mathcal{S}(\mathcal{G}_\eta,c_\walkl)$ in order to to verify \eqref{eq_theo_sum}.
 Thus to be able to show stability of a  DDS~\eqref{eq_def_dds} for a preferably large sampling period $h$, it remains to find the parameters $\gamma_i,L_i,\epsilonm_i$ and $\lvar _i$ for all $i\in\left\lbrace 1,\dots,w(\eta)\right\rbrace$, such that the conditions of Theorem~\ref{theo_stab} hold. For a given $h$, the least conservative parameters are given for each $i\in\left\lbrace 1,\dots,w(\eta)\right\rbrace$ by the solution of \change{the optimization problem}{}
\begin{align}
\begin{split}
\label{eq_opt_prob}
	\left(\gamma_i,L_i,\epsilonm_i, \lvar _i\right) =&  \underset{\gamma_i,L_i,\epsilonm_i \lvar _i}{\arg\inf } \left\lbrace\max\left\lbrace-\epsilonm_i,2(L_i-\lvar _i) \right\rbrace\right\rbrace  \\
	s.t.~ &Ass.~\ref{asum_hybrid_lyap}~\text{\normalfont holds}  \\	 
	&T_{\max}(\gamma_i,\lvar _i) > hi.
	\end{split}
\end{align}
Solving this \change{}{optimization} problem is in general not an easy task, since the solution has to satisfy Assumption~\ref{asum_hybrid_lyap}.

\change{Fortunately, it is not neccesary to find the solution to the optimization problem. It suffices to find a parameter set which is feasible for the constraints of \eqref{eq_opt_prob} in order to use Theorem~\ref{theo_stab}.}{Fortunately, it suffices to find a feasible parameter set for the constraints of \eqref{eq_opt_prob}, instead of a solution to \eqref{eq_opt_prob}, in order to use Theorem~\ref{theo_stab}.} If \eqref{eq_theo_sum} holds for that suboptimal parameter set, then we can conclude asymptotic stability for the DDS~\eqref{eq_def_dds}. 
We will demonstrate in Section~\ref{sec_ex} for an example system from literature how a (possibly suboptimal) but feasible set of parameters can be found 
with SOSTOOLS (\cite{sostools}), such that Theorem~\ref{theo_stab} yields a significantly less conservative stability condition than the \naive approach.
 (or equivalently than using Theorem~\ref{theo_stab} with a constraint of the type $\overline{\binomangle{m-n+1}{*}}$).

\begin{rema}
	The recent results from \cite[Chapter 8]{heijmans2019exploring}, can be interpreted as an alternative approach to deal with additional knowledge about dropout sequences. Therein, the worst case MASP has been significantly prolongated for systems with distributed sensors in comparison to the MASP from \cite{carnevale2007lyapunov} as long as the average sampling time is small enough. However, the results from \cite[Chapter 8]{heijmans2019exploring} are only \change{advantageous}{benificial} for systems with distributed sensors that are not sampled simultaneously and do not improve the MASP for the sampled-data setup that is considered in this paper in comparison to the maximum number of dropouts approach with $T_{\max}(\gamma,L)$ according to \cite{nesic2009explicit}.
\end{rema}
\section{Example}
\label{sec_ex}
\change{In this section}{Next}, we apply Theorem~\ref{theo_stab} to an example \change{system}{}from literature and  demonstrate that \change{Theorem~\ref{theo_stab}}{it} yields a less conservative stability condition than using the \naive approach, when the dropout sequence satisfies a WHRT constraint. 
We consider the unstable example system
\begin{equation}
\dot{x} = d_2x^2-x^3+u,
\end{equation}
that was also used \change{as example system }{}in \cite{nesic2009explicit}, and the constraint $\eta = \binom{17}{20}$ with $w(\eta) = 3$. The Graph $\mathcal{G}_\eta$ can be constructed using Algorithm~1 from \cite{linsenmayer2019linear}. We choose $d_2 = 1$, $\kappa(x) = -2x$, $V(x) = \frac{1}{2}x^4-\frac{2}{3}x^3+2x^2$ and $W(e) = \norm{e}$ and obtain thus $f(x,e) = -2x+x^2-x^3+2e$ and $g(x,e) = -f(x,e)$. We compute 
\begin{align*}
	\left\langle \frac{\partial}{\partial e} W^2(e), g(x,e) \right\rangle =& 2 \norm{e}\mathrm{sign}(e) \left(g(x,e) \right)\\
	\leq& 2 W(e) \left( 2 \norm{e} + \norm{2x-x^2+x^3}\right)
\end{align*}
and observe thus that $L = 2$ and $H(x) = \norm{2x-x^2+x^3}$. \change{For the considered example, feasible}{Feasible} value pairs for $\epsilon$ and $\gamma$, for which \eqref{eq_v_desc_hybrid} holds, can be found with SOSTOOLS (\cite{sostools}). The largest value for $T_{\max}(\gamma,L)$ for $\epsilon > 0$ is achieved for $\gamma = 2$. Thus, for the \naive approach, the best achievable bound on the MASP is $h < \frac{T_{\max}(2,2)}{4}=$~\SI{0.125}{\second}. Further parameter sets, for which Assumption~\ref{asum_hybrid_lyap} holds are given in Table~\ref{tab_param}.
\begin{table}[tb]
	\centering	
	\caption{Feasible parameters for \eqref{eq_opt_prob}\change{, computed using SOSTOOLS}{}.}
	\begin{tabular}{|c|c|c|c|c|c|c|c|}
		\hline
		$i$  & ${\epsilon}_i$&${\gamma}_i$&$\lvar _i$&$L_i$&$T_{\max}(\gamma_i,\lvar _i)$ \\ \hline
		1&1.5&5.77&2.75&2&\SI{0.211}{\second}\\ \hline
		2&0.5&2.38&2.25&2&\SI{0.428}{\second}\\ \hline
		3&-2&2.00&1&2&\SI{0.605}{\second}\\ \hline
		4&-4&2.00&0.001&2&\SI{0.787}{\second}\\ \hline
	\end{tabular}

	\label{tab_param}
\end{table}
For $h \leq$~\SI{0.195}{\second}, these parameters are feasible for the constraints of \eqref{eq_opt_prob}  and are thus suitable to use Theorem~\ref{theo_stab}, even tough they are not the (optimal) solution to \eqref{eq_opt_prob}. 

With the WHRT graph $\mathcal{G}_\eta$, \eqref{eq_theo_sum} can be verified automatically for the parameters from Table~\ref{tab_param} and all walks from $\mathcal{S}(\mathcal{G}_\eta,c_\walkl)$. We observe, that \eqref{eq_theo_sum} holds for $c_\walkl = 20$. Thus, we can use Theorem~\ref{theo_stab} to show asymptotic stability for the example system and $h =$~\SI{ 0.195}{\second}\change{. This means, that for the considered example, stability can be guaranteed with Theorem~\ref{theo_stab}}{, i.e.,} for a sampling period that is $1.56$ times larger than the maximum sampling period for which stability can be guaranteed with the \naive approach. 
\section{Conclusion}
\label{sec_conclusion}
In this paper, we have discussed the stability analysis of nonlinear sampled-data systems that run in open-loop from time to time due to failures in the feedback channel. WHRT constraints were used as precise dropout description. The overall approach is based on the emulation technique from \cite{nesic2009explicit}, and is in general less conservative than considering only the maximum number of successive dropouts. The efficiency of the proposed approach has been illustrated for an example system from literature\change{ for which stability could be guaranteed for a significantly prolongated sampling period in comparison to the \naive approach}{}.

\bibliography{../../../../../Literatur/literature}
\normalsize
\arxivchange{}{\appendix
\section{}
\textbf{Proof of Proposition~\ref{prob_michel}.}
\label{append_a}
(This proof follows directly from the proof of Theorem 6.4.6 from \cite{michel2015stability}).
We have to show that $\forall \beta> 0, \exists \delta(\beta) > 0, s.t.$
\begin{align}
 \norm{\xi(0)} \leq \delta(\beta)
 \Rightarrow \norm{\xi(t)} \leq \beta~ \forall t \geq 0 \label{eq_eps_del}
\end{align}and that
\begin{equation}
\label{eq_asym_conv}
\norm{\xi(t)} \rightarrow 0~ \text{\normalfont as }t \rightarrow \infty
\end{equation}  for all $\xi(0) \in \mathbb{R}^{2n}$, to proof asymptotic stability. Subsequently, we use the abbreviation $\psi_z \defeq V_n(\xi(\tilde{\tau}_z))$. From \eqref{eq_non_mon_desc2}, we observe that $\psi_z \leq \psi_0$ for all $z \geq 0$. Using \eqref{eq_non_mon_dec4} and \eqref{eq_non_mon_desc1}, we obtain for $\tilde{\tau}_z \leq t < \tilde{\tau}_{z+1}$ that 
\begin{equation}
\label{eq_cont_z_bound}
\norm{\xi(t)} \leq \alpha_1^{-1}\left(\alpha_3 \left(\psi_z\right)\right) 
\end{equation}
and thus $\norm{\xi(t)} \leq \alpha_1^{-1}\left(\alpha_3 \left(\psi_0\right)\right)$ for all $t \geq 0$.
Hence, $\norm{\xi(t)} \leq \beta$ if $\norm{\xi_0} \leq  \alpha_2^{-1} \left(\alpha_3^{-1} \left(\alpha_1 \left( \beta \right)\right)\right)\defeq \delta(\beta)$ and thus \eqref{eq_eps_del} is satisfied. To show \eqref{eq_asym_conv}, we consider again \eqref{eq_non_mon_desc2} and observe by using \eqref{eq_non_mon_desc2} recursively that $\psi_{z+1} - \psi_{0} \leq - (\tilde{\tau}_{z+1} - 0) \alpha_4(\psi_z)$. With \eqref{eq_non_mon_dec3}, this yields $\psi_z \leq \alpha_4^{-1} \left( \frac{\psi_0}{(z+1)\underline{h}}\right)$. Condition \eqref{eq_asym_conv} follows directly due to \eqref{eq_cont_z_bound} and since $\tilde{\tau}_{z+1} - \tilde{\tau}_z \leq \overline{h}$.
\hfill\hfill \qed

\section{}
\label{append_b}
	\textbf{Proof of Proposition~\ref{prop_hybrid}}.	
	In this proof, we will construct a Lyapunov like function based on $V(x)$ and $W(e)$ and show using Assumption~\ref{asum_hybrid_lyap} that the amount of increase or decrease of this function for times between $\tau_z$ and $\tau_{z+1}$ is bounded explicitly depending on the time span between $\tau_z$ and $\tau_{z+1}$. From this function we can draw conclusions for $V(x(t))$ and $W(e(t))$ for $\tau_z < t \leq \tau_{z+1}$. First, we state some preliminaries.
	Let $\phi : [0,\tilde{T}_{\max}] \rightarrow \mathbb{R}$ be the solution of
	\begin{equation}
	\dot{\phi} = -2\lvar \phi-\gamma(\phi^2+1),~ \phi(0) = \lambda^{-1}
	\end{equation}
	for some $\lambda \in \left( 0,1\right)$. Recall from \cite{carnevale2007lyapunov} that $\phi(\tau) \in \left[\lambda, \lambda^{-1}\right]$ for all $\tau \in \left[0, \tilde{T}_{\max} \right]$, where $\tilde{T}_{\max} = \tilde{T}_{\max}(\lambda,\gamma,\lvar )$ with
	\begin{footnotesize}		
		\begin{equation*}
		\tilde{T}_{\max}(\lambda,\gamma,\lvar ) \defeq \begin{cases}\vspace{1mm}
		\frac{1}{\lvar r} \mathrm{arctan}\left(\frac{r(1-\lambda)}{2 \frac{\lambda}{1+\lambda} \left(\frac{\gamma}{\lvar }-1\right)+1+\lambda}\right) & \gamma > \lvar \\ \vspace{1mm}
		\frac{1}{\lvar } \frac{1-\lambda}{1+\lambda} & \gamma = \lvar \\
		\frac{1}{\lvar r} \mathrm{arctanh}\left(\frac{r(1-\lambda)}{2 \frac{\lambda}{1+\lambda} \left(\frac{\gamma}{\lvar }-1\right)+1+\lambda}\right) &\gamma < \lvar 
		\end{cases}
		\end{equation*}
	\end{footnotesize}
	and  $r$ defined as in \eqref{eq_def_r}.

	Now, we consider \eqref{eq_aux_sys} for arbitrary $z \in \mathbb{N}_0$. For each $t_{\max}$ with  $t_{\max} - \tau_z < T_{\max} (\gamma,\lvar )$, there is a $\lambda \in \left(0,1\right)$ such that $t_{\max}  - \tau_z = \tilde{T}_{\max}(\lambda,\gamma,\lvar )$. We introduce the additional state ${\mu}$, with $\mu(\tau_{z})  =  0$ and $\dot{\mu} = 1$ for $\tau_z \leq t < t_{\max}$.  We will subsequently use  $\tilde{\zeta} \defeq \left[\tilde{x}^\top,\tilde{e}^\top,\mu\right]^\top$ and $F(\tilde{\zeta})\defeq \left[f(\tilde{x},\tilde{e})^\top,g(\tilde{x},\tilde{e})^\top,1\right]^\top$. Note that a unique solution to the differential equation $\dot{\tilde{\zeta}}(t) = F(\tilde{\zeta}(t))$ exists for some time after $t = \tau_z$, if we fix $\tilde{x}(\tau_z) = x(\tau_z)$ and $\tilde{e}(\tau_z) = 0$,
	  due to the continuous differentiability of $F(\tilde{\zeta}(t))$.  We define 
	\begin{equation}
	U(\tilde{\zeta}(t)) \defeq V(\tilde{x}(t)) + \gamma \phi(\mu(t))W^2(\tilde{e}(t))
	\end{equation}
	for which we can show (cf. \cite{nesic2009explicit}) that
	\begin{equation}
	\label{eq_U_lyap}
		\underline{\alpha}_U\left(\norm{\left[x^\top,e^\top\right]^\top}\right) \leq U(\tilde{\zeta}) \leq \overline{\alpha}_U\left(\norm{\left[x^\top,e^\top\right]^\top}\right)
	\end{equation}
	 for some $\underline{\alpha}_U, \overline{\alpha}_U \in \mathcal{K}_\infty$. We observe due to Assumption~\ref{asum_hybrid_lyap} that for $\tau_z \leq 
	 t < t_{\max}$ and all $x\in\mathbb{R}^{n_x}$ and $e \in \mathbb{R}^{n_x}$
	\begin{align*}
	\frac{d}{dt} U(\tilde{\zeta}(t)) =&\left\langle \nabla U(\tilde{\zeta}(t)), F(\tilde{\zeta}(t)) \right\rangle\\
	\leq&-\epsilonm V(\tilde{x}(t)) -H^2(\tilde{x}(t)) + \gamma^2 W^2(\tilde{e}(t)) \\
	&+2\gamma\phi(\mu(t))W(\tilde{e}(t))(L W(\tilde{e}(t)) + H(\tilde{x}(t)))\\
	& - \gamma W^2(\tilde{e}(t)) (2\lvar \phi(\mu(t))+\gamma(\phi^2(t)+1))\\
	\leq&-\epsilonm V(\tilde{x}(t)) -(H(\tilde{x}(t)) -\gamma\phi(\mu(t))W(\tilde{e}(t)))^2\\ 
	&+ 2\gamma\phi(\mu(t))W^2(\tilde{e}(t))(L-\lvar )\\
	\leq& -\epsilonm V(\tilde{x}(t)) + 2\gamma\phi(\mu(t))W^2(\tilde{e}(t))(L-\lvar ).
	\end{align*} 
	
	Thus, we have that for $\tau_z \leq	t < t_{\max}$
	\begin{equation}
	\label{eq_U_desc}
	\frac{d}{dt} U(\tilde{\zeta}(t)) \leq \max \left\lbrace -\epsilonm, 2(L - \lvar ) \right\rbrace U(\tilde{\zeta}(t)).
	\end{equation} 
	This implies with \eqref{eq_U_lyap} and Lyapunov like arguments, that a solution to \eqref{eq_aux_sys} exists  for $\tau_z \leq t < t_{\max}$. We can choose $t_{\max}$ such, that  $\tau_{z+1} < t_{\max} < \tau_z + T_{\max} (\gamma,\lvar )$. Hence, a unique solution to \eqref{eq_aux_sys} exists  for $\tau_z \leq t \leq \tau_{z+1}$.
	
	Now we consider \eqref{eq_def_dds} and $\zeta \defeq \left[x^\top,e^\top,\mu \right]^\top$ with $\dot{\zeta} = F(\zeta)\defeq \left[f(x,e)^\top,g(x,e)^\top,1 \right]^\top$. We observe that $x(t) = \tilde{x}(t)$ and $e(t) = \tilde{e}(t)$ for $\tau_z \leq t < \tau_{z+1}$. Hence, $U(\zeta(t)) = U(\tilde{\zeta}(t))$  for $\tau_z \leq t < \tau_{z+1}$.
	We note that 
	\begin{equation}
	\label{eq_U_start}
	U(\zeta(\tau_{z})) = V(x(\tau_{z})) + \gamma \phi(0) W^2(e(\tau_{z})) \overset{e(\tau_z = 0)}{=} V(x(\tau_{z})).
	\end{equation}
	 Due to the comparison Lemma (cf. \cite[p. 102]{khalil2002nonlinear}) and \eqref{eq_U_desc}, we obtain moreover for $\tau_z \leq t < {\tau_{z+1}}$ that
	\begin{align}
	U(\zeta(t)) \leq& \exp\left( \max \left\lbrace -\epsilonm, 2(L - \lvar ) \right\rbrace (t-\tau_z)\right)U(\zeta(\tau_z))\nonumber \\
	\leq& \exp\left( \max \left\lbrace -\epsilonm, 2(L - \lvar ) \right\rbrace (t-\tau_z)\right)V(x(\tau_z)). \nonumber
	\end{align}
	Hence, \eqref{eq_prop_hybrid2} holds with 
	\begin{equation*}
	k_1 =  \max\left\lbrace\exp\left( \max \left\lbrace -\epsilonm, 2(L - \lvar ) \right\rbrace (\tau_{z+1} - \tau_z)\right),1\right\rbrace.
	\end{equation*}
	Furthermore, we have that
	\begin{align}
	&V(x(\tau_{z+1}))\nonumber\\ 
	=&  V(x(\tau_{z+1}^-)) \leq  V(x(\tau_{z+1}^-)) + \gamma \phi(\tau_{z+1}^--\tau_z) W^2(e(\tau_{z+1}^-))\nonumber \\
	=&  \nonumber  U(\zeta(\tau_{z+1}^-)) \nonumber	\\
	\leq&   \exp\left( \max \left\lbrace -\epsilonm, 2(L - \lvar ) \right\rbrace (\tau_{z+1}^--\tau_z)\right)V(x(\tau_z)) \nonumber \\
	=& \exp\left( \max \left\lbrace -\epsilonm, 2(L - \lvar ) \right\rbrace (\tau_{z+1}-\tau_z)\right)V(x(\tau_z)) \nonumber
	\end{align}

	and hence \eqref{eq_prop_hybrid1} holds.	
	Additionally, we have that 
	\begin{align*}
	&W^2(e(t))\\
	\leq& \frac{1}{\gamma \phi(\mu(t))} \exp\left( \max \left\lbrace -\epsilonm, 2(L - \lvar ) \right\rbrace (t-\tau_z)\right)V(x(\tau_z))\\
	\leq& k_2 V(x(\tau_z))
	\end{align*}
	with 
	\begin{equation*}
	k_2 = \frac{1}{\gamma\lambda} \max\left\lbrace \exp\left( \max \left\lbrace -\epsilonm, 2(L - \lvar ) \right\rbrace (\tau_{z+1}-\tau_z)\right),1\right\rbrace.
	\end{equation*}
	From \eqref{eq_w_bound}, it follows  that 
	\begin{equation*}
	\norm{e(t)} \leq \underline{\alpha}_W^{-1}\left(\sqrt{W^2(e)}\right)  \leq \underline{\alpha}_W^{-1}\left(\sqrt{k_2V(x(\tau_z))}\right) .
	\end{equation*}
	Since the inverse of a class $\mathcal{K}$ function is of class $\mathcal{K}$ and concatenations of class $\mathcal{K}$ functions are of class $\mathcal{K}$, $\underline{\alpha}_W^{-1}\left(\sqrt{k_2(.)}\right)$ is a class $\mathcal{K}$ function.
	Thus, \eqref{eq_prop_hybrid3} holds with
	$\alpha_{w2}(.) =\underline{\alpha}_W^{-1}\left(\sqrt{k_2 (.)}\right)$. 
	\hfill\hfill \qed

\section{}
\label{append_c}
\textbf{Proof of Theorem~\ref{theo_stab}.}
	We show now
	  that the conditions of Proposition~\ref{prob_michel} hold for the DDS~\eqref{eq_def_dds} and $V_n(\xi) = V(x)+\norm{e}$, such that we can conclude asymptotic stability of the origin of \eqref{eq_def_dds}. 
	  
	  First, we note that \eqref{eq_non_mon_dec4} holds for $V_n$ due to Assumption~\ref{as_cont_fb}.	
	The next step is to define $\tilde{\mathcal{T}}$. We chose $\tilde{\tau}_0 = \tau_0 = 0$. 
 The rest of the sequence $(\tilde{\tau}_z)_{z \in \mathbb{N}_0}$ for Proposition~\ref{prob_michel} is defined iteratively as follows. For arbitrary fixed $z\in\mathbb{N}_0$ and $\tilde{\tau}_z$, we choose  $\pi \in \mathbb{N}_0$ such that $\tau_\pi = \tilde{\tau}_z$. 	Recall that $(\tau_z)_{z\in\mathbb{N}_0}$ is the sequence of successful transmissions.  Since $\nu \vdash \eta$, and since $\mathcal{G}_\eta$ is a WHRT graph for $\eta$, there is a\footnote{If there are several walks in $\mathcal{S}(\mathcal{G}_\eta,c_\walkl)$ satisfying the condition, then we can choose one of them arbitrarily.} $P\in \mathcal{S}(\mathcal{G}_\eta,c_\walkl)$ such that $\tau_{\pi+j}-\tau_{\pi+j-1} = l_{\mathbb{I}_j(P)}h$ for all $j \in \left\lbrace 1,\dots,n_P \right\rbrace$. 
	We then define $\tilde{\tau}_{z+1} = \tau_{\pi+n_P}$. The cost of $P$ is bounded by $c_\walkl+w(\eta)$.
	 Thus, \eqref{eq_non_mon_dec3} holds with $\underline{h} = h$ and $\overline{h} = h (c_\walkl+w(\eta))$ for the considered choice of $\tilde{\mathcal{T}}.$
	 	
	Next, we show that \eqref{eq_non_mon_desc1} and \eqref{eq_non_mon_desc2} hold for $\tilde{\tau}_z$ and $\tilde{\tau}_{z+1}$ according to the above definition for each $z \in \mathbb{N}_0$, by exploiting conditions \eqref{eq_sample_bound} and \eqref{eq_theo_sum}. 
	Note that $l_{\mathbb{I}_j(P)} \in \left\lbrace 1, \dots,w(\eta) +1\right\rbrace$ for all $j \in \left\lbrace 1,\dots,n_P \right\rbrace$. Since $\tau_{\pi+j}-\tau_{\pi+j-1} = hl_{\mathbb{I}_j(P)} < T_{\max}(\gamma_{l_{\mathbb{I}_j(P)}},\lvar _{l_{\mathbb{I}_j(P)}})$ and because a new input was received at time $\tau_{\pi+j-1}$, we know due to Proposition~\ref{prop_hybrid} that 
		\begin{align}
		\nonumber V(x(\tau_{\pi+j})) 
		\leq& \exp\left(\max\left\lbrace -\epsilon_{l_{\mathbb{I}_j(P)}},\right.\right. \\ 
		&\left.\left.2(L_{l_{\mathbb{I}_j(P)}}-\lvar _{l_{\mathbb{I}_j(P)}}) \right\rbrace hl_{\mathbb{I}_j(P)}\right) V(x(\tau_{\pi+j-1})). \label{eq_desc_j}
		\end{align}
		Furthermore, Proposition~\ref{prop_hybrid} implies that 	
		\begin{equation}
		\label{eq_x_desc}
		V(x(t)) \leq k_{1} V(x(\tau_{\pi+j-1}))
		\end{equation}
		and
		\begin{equation}
		\label{eq_e_desc}
		\norm{e(t) }\leq \alpha_{w2}(V(x(\tau_{\pi+j-1})))
		\end{equation}
		hold for some $k_{1}\in \mathbb{R}_{>0}$, $\alpha_{w2}\in\mathcal{K}$ and $\tau_{\pi+j-1} \leq t < \tau_{\pi+j}$ which will be used later in the proof. 
		
		Now, we can use \eqref{eq_desc_j} iteratively starting at $\tau_\pi = \tilde{\tau}_z$ and obtain that
		\begin{align}
		\nonumber V(x(\tau_{\pi+j})) 
		\leq& \exp \left( \sum_{\theta = 1}^{j}\left(\max\left\lbrace -\epsilon_{l_{\mathbb{I}_\theta(P)}}, \vphantom{2(L_{l_{\mathbb{I}_\theta(P)}}-\lvar _{l_{\mathbb{I}_\theta(P)}}}\right.\right.\right.\\
		&\left.\left. \left. 2(L_{l_{\mathbb{I}_\theta(P)}}-\lvar _{l_{\mathbb{I}_\theta(P)}}) \right\rbrace hl_{\mathbb{I}_\theta(P)}\right) \vphantom{\sum_{\theta = 1}^{j}}\right) V(x(\tau_{\pi})) \label{eq_v_sum}.
		\end{align}
		Thus, we observe that 
		\begin{equation*}
			V(x(\tilde{\tau}_{z+1})) - V(x(\tilde{\tau}_{z})) \leq -\left(\tilde{\tau}_{z+1}-\tilde{\tau}_{z}\right) k_3 V(x(\tilde{\tau}_{z}))
		\end{equation*} 
		holds for all $z\in\mathbb{N}_0$ with
		\begin{align*}
			k_3 =& \frac{1}{\bar{h}} \left(1-\exp\left(\underset{P \in \mathcal{S}(\mathcal{G}_\eta,c_\walkl)}{\max}\left\lbrace\vphantom{\sum_{\theta = 1}^{n_{P}} hl_{p_\theta,P}}\right.\right.\right.\\
				 &\left.\left.\left. \sum_{\theta = 1}^{n_{P}} hl_{\mathbb{I}_\theta(P)} \max \left\lbrace -\epsilon_{l_{\mathbb{I}_\theta(P)}},2(L_{l_{\mathbb{I}_\theta(P)}}-\lvar _{l_{\mathbb{I}_\theta(P)}})\right\rbrace \right\rbrace \right)\right).
		\end{align*}
		We can conclude from \eqref{eq_theo_sum} that $k_3 > 0$, and thus 
		\begin{equation*}
		\frac{1}{\tilde{\tau}_{z+1} - \tilde{\tau}_z} \left[ V(x(\tilde{\tau}_{z+1})) - V(x(\tilde{\tau}_z))\right] \leq -\alpha_4 (V(x(\tilde{\tau}_z)))
		\end{equation*}
		holds for all $z\in\mathbb{N}_0$ with $\alpha_4(r) = k_3r$. Since $\norm{e(\tilde{\tau}_z)} = 0$ for all $z\in\mathbb{N}_0$ due to the structure of \eqref{eq_def_dds}, this implies that \eqref{eq_non_mon_desc2} holds. Now, it remains to show that \eqref{eq_non_mon_desc1} holds. We use \eqref{eq_x_desc} and \eqref{eq_e_desc} and obtain
		\begin{align}
			V_n(\xi(t)) =& V(x(t)) + \norm{e(t)}\nonumber \\
			\leq& k_{1} V(x(\tau_{\pi+j-1})) + \alpha_{w2}(V(x(\tau_{\pi+j-1})))\nonumber \\
			\leq& \alpha_{5}(V(x(\tau_{\pi+j-1})) \label{eq_v_dec}
		\end{align}
		for  $\tau_{\pi+j-1} \leq t < \tau_{\pi+j}$, where 
		\begin{equation*}
			\alpha_{5}(r) =  2\max\left(k_{1}r, \alpha_{w2}(r) \right) \in \mathcal{K}.
		\end{equation*}
		From \eqref{eq_v_sum}, we can conclude that $V(x(\tau_{\pi+j-1})\leq \linebreak k_{4} V(x(\tilde{\tau}_{z})) $ for some $k_{4} \in \mathbb{R}_{>0}$ and for all $j\in\left\lbrace 1,\dots,n_P\right\rbrace$. Together with \eqref{eq_v_dec}, this implies that \eqref{eq_non_mon_desc1} holds for all $z \in \mathbb{N}_0$ and thus asymptotic stability follows for all $\nu\vdash \eta$. The fact that all sequences that satisfy $\eta'$ also satisfy $\eta$ concludes the proof. \hfill\hfill \qed
}

\end{document}